\documentclass[aps,preprintnumbers,prl,twocolumn,groupedaddress]{revtex4}

\usepackage{graphicx}

\def\be{\begin{equation}}
\def\ee{\end{equation}}
\def\bea{\begin{eqnarray}}
\def\eea{\end{eqnarray}}

\def\bbuildrel#1_#2^#3{\mathrel{\mathop{\kern 0pt#1}\limits_{#2}^{#3}}}
\def\slash#1{\setbox0=\hbox{$#1$}#1\hskip-\wd0\dimen0=5pt\advance
       \dimen0 by-\ht0\advance\dimen0 by\dp0\lower0.5\dimen0\hbox
         to\wd0{\hss\sl/\/\hss}}

\newcommand{\gae}{\lower 2pt \hbox{$\, \buildrel {\scriptstyle >}\over {\scriptstyle
\sim}\,$}}
\newcommand{\lae}{\lower 2pt \hbox{$\, \buildrel {\scriptstyle <}\over {\scriptstyle
\sim}\,$}}

\begin{document}

\title{Possible Indications of New Physics in $B_d$-mixing and in $\sin (2\beta)$ Determinations}

\author{Enrico Lunghi}
\email{lunghi@fnal.gov}
\affiliation{Fermi National Accelerator Laboratory, P.O. Box 500 , Batavia, IL, 60510-0500, USA}

\author{Amarjit Soni}
\email{soni@quark.phy.bnl.gov}
\affiliation{Physics Department, Brookhaven National Laboratory, Upton, New York, 11973, USA}

\preprint{BNL-HET-08/8}
\preprint{FERMILAB-PUB-08-073-T}

\begin{abstract}

Using the hadronic matrix elements from the lattice, $B_K$ and $\xi_s$, involving only the 4-quark operators for $\Delta$ flavor $=$ 2 Hamiltonian relevant for $K-\bar K$, $B_d-\bar B_d$ and $B_s - \bar B_s$ mixing, along with $V_{cb}$, we deduce a non-trivial constraint on the SM, $\sin (2 \beta) = 0.87\pm 0.09$. This deviates from direct experimental measurements via the tree process, $b \to c \bar c s$ as well as the one via the penguin-loop $b \to s$ decays by around 2.1 and 2.7 $\sigma$ respectively. If these deviations are confirmed they would imply the presence of new physics rather pervasively in both $B_d - \bar B_d$ ({\it i.e} very likely in $b \to d$) as well as in $b \to s$ transitions requiring a beyond the SM CP-odd phase. Consequently, improvements in the relevant lattice calculations should be given a high priority.

\end{abstract}

\maketitle

In the past several years spectacular performance of the B-factories has led to a milestone in our understanding of CP violation phenomena. Results from the B-factories, along with Standard Model (SM) predictions, based in large part on lattice calculations of relevant weak matrix elements, provided a striking confirmation of the Cabibbo-Kobayashi-Maskawa (CKM)~\cite{Cabibbo:1963yz,Kobayashi:1973fv} paradigm~\cite{Browder:2008em}. For the first time it was experimentally established that the single (CP-odd) phase of the KM ansatz~\cite{Kobayashi:1973fv} is able to account to an accuracy of about 15\% the observed O(1) CP asymmetry seen in the gold-plated, $B \to \psi K_S$ type of modes as well as the minuscule ({\it i.e.} O($10^{-3}$)) CP asymmetry seen in $K_L$ decays long ago~\cite{Christenson:1964fg}! While this is a remarkable success of the CKM picture, we should note that there are essentially compelling theoretical expectations that beyond the Standard Model (BSM) CP-odd phase(s) must also exist in nature. Finding evidence for these is clearly an important challenge for Particle Physics. 

One strategy to search for these new phase(s) is to look for deviations from the SM in its precise  description of the asymmetry as seen in (say) $B \to \psi K_s$ decays. Since in this case one is clearly looking for effects that are small in comparison to the dominant contribution of the SM, precision in theory and in experiment becomes highly desirable. Another class of searches are amongst those channels wherein the SM predicts vanishing asymmetries; these belong to a class of null tests~\cite{Gershon:2006mt,Browder:2008em} for the SM. Sizable CP asymmetries in such channels would then have to be ascribed to BSM sources. In this work, we will show that using clean input from lattice calculations leads to extremely interesting indications of beyond the SM CP-odd phases(s) in both of the aforementioned class of tests. Furthermore, there are good reasons to expect that these lattice calculations can be improved further to provide even more stringent tests.

In this work, using primarily the input from the lattice for the non-perturbative matrix elements of the four-quark operator of the $\Delta$ flavor (F) $=$ 2, SM Hamiltonian, we show that the value of $\sin (2 \beta)$ that we thus determine, with this clean input, exhibits some  deviations from that measured directly in B-factory experiments. These discrepancies are potentially of crucial importance as they signify the possible presence of physics beyond the Standard Model in $B_d-\bar B_d$ mixing, as well as in $b \to s$ penguin transitions. These observations lead  to the important conclusion that the effects of physics beyond the SM are possibly making their presence felt rather pervasively both in $b \to d$ as well as in $b \to s$ short distance penguin transitions. These deviations are specially important as they are obtained without making any use of $V_{ub}$, which has been very problematic in recent years as the value deduced from inclusive methods disagrees appreciably with that deduced from the exclusive approaches~\cite{Yao:2006px}.

To reiterate, we show that, along with $V_{cb}$, only the hadronic matrix elements of one type of 4-quark operator: $[\bar h \gamma_\mu (1- \gamma_5) q]^2$, with $h=s$ ($q=d$) and $h=b$ ($q=s$ and $d$) are needed for providing a non-trivial constraint on the value of $\sin (2 \beta)$ in the SM. Note that for all three of the relevant mixing-matrix elements, for $K$, $B_d$ and $B_s$ mesons, in fact the strange and the $b$-quarks can be considered very heavy 
compared to the light $u$ and $d$ quarks, explicitly demonstrated recently by the RBC-UKQCD collaborations' determination of $B_K$~\cite{Antonio:2007pb}. Thereby only $SU(2) \times SU(2)$ (rather than $SU(3) \times SU(3)$) chiral perturbation theory need be used for chiral extrapolations from the light quarks on the lattice to their physical value resulting in significantly improved accuracy. Furthermore, it is important to note that from B-mixings in fact we 
only use the $SU(3)$ breaking ratio of mass differences, $\Delta M_{B_s}/ \Delta M_{B_d}$~\cite{Bernard:1998dg}. The prognosis for steady improvement in the accuracy of these matrix elements therefore seems to be quite good.

We now present the analysis based on the thinking outlined above. Fig.~\ref{fig:utfit_novub} shows the region of the $(\bar \rho,\bar \eta)$ plane allowed by
$|V_{cb}|$, $\Delta M_{B_s}/\Delta M_{B_d}$ and $\varepsilon_K$. No use of $V_{ub}$ is made here; this is an important difference from our previous work~\cite{Lunghi:2007ak}. This is done to alleviate any concerns that the problems afflicting $V_{ub}$~\cite{Yao:2006px,Neubert:2008cp}, that we alluded to above, may be unduly effecting the results. The key inputs being used are as follows:
\begin{figure}
\includegraphics[width= 0.95 \linewidth]{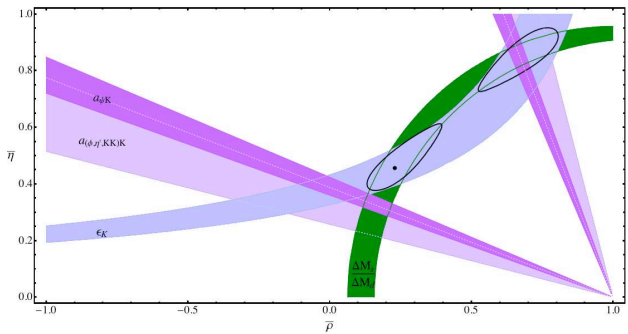}
 \caption{Unitarity triangle fit in the SM. All constraints are imposed at the 68\% C.L.. The solid contour is obtained using the constraints from $\varepsilon_K$ and $\Delta M_{B_s}/\Delta M_{B_d}$. The regions allowed by $a_{\psi K}$ and $a_{(\phi+\eta^\prime + 2 K_s)K_s}$ are superimposed.\label{fig:utfit_novub}}
\end{figure}

\begin{itemize}
\item The $K-\bar K$ mixing matrix element conventionally parametrized as $\hat B_K$~\cite{Antonio:2007pb} needed from the lattice to make use of the indirect CP-violation parameter, $\epsilon_K$ from $K_L \to \pi \pi$,
\bea
 \hat B_K = 0.720 \pm 0.013 \pm 0.037
\eea
\item The SU(3) breaking ratio, $\xi_s$, from the lattice, needed in
conjunction with $\Delta m_{B_s}/\Delta m_{B_d}$, which is the ratio of
the $B_s$ and $B_d$ mass differences~\cite{Tantalo:2007ai,Gamiz:2007tg,Becirevic:2003hf},
\bea
\xi_s  =  \frac{f_{B_s} \sqrt{\hat B_s}}{f_{B_d} \sqrt{\hat B_d}}  &=&
1.20 \pm 0.06
\eea
\item As far as $V_{cb}$, which is another input that is necessary, we note that there is a $\approx 2 \sigma$ tension between the extraction of $|V_{cb}|$ from inclusive and exclusive decays~\cite{Buchmuller:2005zv,Laiho:2007pn}:
\bea
\left|V_{cb}\right| \times 10^3= \cases{
                41.7 \pm 0.4 \pm 0.6 & inclusive\cr
		                38.7 \pm 0.7 \pm 0.9 & exclusive \cr}
\eea
Therefore, in the numerics we use the weighted average of these two determinations:
\bea
\left| V_{cb} \right| = (40.8 \pm 0.6 ) \times 10^{-3} \; .
\eea
\end{itemize}

Thus from Fig.~\ref{fig:utfit_novub} it is clear that even without the inclusion of $|V_{ub}|$, the prediction for $\sin (2\beta)$ deviates from the experimental
determinations summarized in Table~\ref{tab:sin2beta}. From the chi--squared minimization we find:
\bea
\left[\sin ( 2 \beta) \right]^{\rm prediction}_{{\rm no} \; V_{ub}} =
0.87 \pm 0.09 \; .
\label{sin2beta_novub}
\eea
Our fitting procedure consists in writing a chi-squared that includes all experimental measurements and lattice determinations; this implies that we assume gaussian errors. The input values that we use in the fits are summarized in Table~\ref{tab:utinputs}. The SM expressions for $\Delta M_{B_s}/\Delta M_{B_d}$ and $\varepsilon_K$ can be found, for instance, in Ref.~\cite{Buras:1998raa}. Following Ref.~\cite{Buras:2008nn}, we include in $\varepsilon_K$ the term proportional to the $I=0$ component of the $K\to\pi\pi$ amplitude, whose contribution is effectively taken into account by the multiplicative factor $\kappa_\varepsilon$. The calculation of the latter is affected by non-perturbative uncertainties and, following the analysis in Ref.~\cite{Anikeev:2001rk,Andriyash:2003ym,Andriyash:2005ax,Buras:2008nn}, we take $\kappa_\varepsilon = 0.92 \pm 0.02$~\cite{Buras-footnote}.
\begin{table}[t]
\begin{center}
\begin{tabular}{|l|}
\hline
$\varepsilon_K = (2.232 \pm 0.007 ) \times 10^{-3}$\\
$\Delta m_{B_s} = (17.77 \pm 0.10 \pm 0.07)\;  {\rm ps}^{-1}$~\cite{Evans:2007hq} \\
$\Delta m_{B_d} = (0.507 \pm 0.005)\; {\rm ps}^{-1}$ \\
$\left| V_{cb} \right| = (40.8 \pm 0.6 ) \times 10^{-3}$ \\
$\hat B_K = 0.720 \pm 0.013 \pm 0.037$~\cite{Antonio:2007pb} \\
$\xi_s  = 1.20 \pm 0.06$ \\
$\lambda = 0.2255  \pm 0.0007$~\cite{Antonelli:2008jg} \\ 
$m_{t, pole} = (170.9 \pm 1.8 ) \; {\rm GeV}$~\cite{:2007bxa} \\
$m_c(m_c) = (1.224 \pm 0.017 \pm 0.054 ) \; {\rm GeV}$~\cite{Hoang:2005zw}\\
$\eta_1 = 1.51 \pm 0.24$~\cite{Herrlich:1993yv} \\
$\eta_2 = 0.5765 \pm 0.0065$~\cite{Buras:1990fn}  \\
$\eta_3 = 0.47 \pm 0.04$~\cite{Herrlich:1995hh}  \\ 
\hline
\end{tabular}
\caption{Inputs that we use in the unitarity triangle fit. When not explicitly stated, we take the inputs from the Particle Data Group~\cite{Yao:2006px}. 
\label{tab:utinputs}}
\end{center}
\end{table}

The constraint in Eq.~(\ref{sin2beta_novub}) deviates by $2.1\sigma$ from $\sin (2\beta)$ extracted from the tree-level decay $B\to J/\psi K_S$ and by $2.7\sigma$ from the average of the three penguin--dominated $B\to (\phi,\eta^\prime, K_S K_S) K_S$ modes (see Table~\ref{tab:sin2beta}). We have chosen to concentrate on the time-dependent CP in these three $b \to s$ penguin-dominated modes (amongst many) as QCD factorization as well as several other approaches show that for these the QCD corrections are very small~\cite{Beneke:2005pu,ccs,bhnr,zw}. In passing, we do want to draw attention, though, to the fact that there are many more $b \to s$ penguin-dominated modes~\cite{Barberio:2007cr}, and curiously in most of  these modes, the central value of the time dependent asymmetry is below the predicted SM value of Eq.~(\ref{sin2beta_novub})~\cite{Lunghi:2007ak} or for that matter the value directly measured via $B \to \psi K_s$, $0.681 \pm 0.025$, indicating again possible problem with the SM description of the $b \to s$ penguin dominated modes. Thus the fact that these data show indications for the need for a BSM-CP-odd phase reported earlier~\cite{Lunghi:2007ak}  does not appear to originate from the use of a faulty input for $V_{ub}$.    

Note that in part this clean  constraint Eq.~(\ref{sin2beta_novub}) has been made possible without any use of $V_{ub}$ due to the significant (almost a factor of three) reduction in errors in lattice determination of $B_K$ achieved in the past few years~\cite{contrast}.   In Table~\ref{tab:sin2beta} we compare the prediction (\ref{sin2beta_novub}) to the various tree and penguin modes. While these penguin modes have been already getting considerable attention in the past $\approx$ 2 years, our analysis makes clear that the low value of $\sin (2 \beta)$ that they yield compared to the SM constraint in Eq.~(\ref{sin2beta_novub}) has nothing to do with the difficulties affecting $V_{ub}$. Furthermore, we find that the difficulties of the SM description of CP violating B-decays may not just be confined to the $b \to s$ penguin modes but in fact may even be there in the ``gold-plated'' $B \to \psi K_s$ mode. Since the latter is a combination of $B_d-\bar B_d$ mixing and the tree decay, $b \to c \bar c s$, it is natural to be more suspicious that BSM physics in $\Delta b=2$, loop process,  may be causing this deviation in the former, though the tree decay may also be in part the cause.   
\begin{figure}
\includegraphics[width= 0.95 \linewidth]{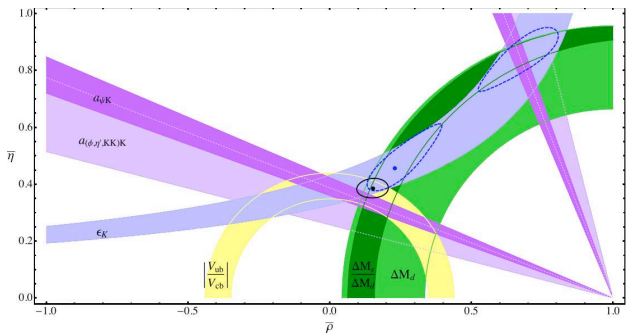}
 \caption{Unitarity triangle fit in the SM. All constraints are imposed at the 68\% C.L.. The solid contour is obtained using the constraints from $\varepsilon_K$, $\Delta M_{B_s}/\Delta M_{B_d}$ and $|V_{ub}/V_{cb}|$. The dashed contour shows the effect of excluding $|V_{ub}/V_{cb}|$ from the fit. 
The regions allowed by $a_{\psi K}$ and $a_{(\phi+\eta^\prime + 2 K_s)K_s}$ are superimposed.\label{fig:utfit}}
\end{figure}

We now turn our attention to $V_{ub}$ to see how it can change the above results. As has recently been  emphasized by  Neubert~\cite{Neubert:2008cp},  
the extraction of $|V_{ub}|$ from inclusive decays is extremely sensitive to the precise determination of the $b$ quark mass. Here we follow Neubert's analysis and do not use $B\to X_s \gamma$ for the extraction of $m_b$. The most recent determinations of $|V_{ub}|$, from inclusive and exclusive works are~\cite{Neubert:2008cp,Dalgic:2006dt,mackenzie-heidelberg}:
\bea
\left|V_{ub}\right| \times 10^4= \cases{
                37.0 \pm 1.5 \pm 2.8 & inclusive \cr
	       35.5 \pm 2.5 \pm 5.0  & HPQCD \cr
	       37.8 \pm 3.0 \pm 3.4 \pm 2.5  & FNAL/MILC \cr}
\eea
The two determinations of $|V_{ub}|$ from exclusive decays are based on the same experimental data and we chose to adopt the Fermilab/MILC~\cite{mackenzie-heidelberg,hpqcd_vub_footnote} determination of the semileptonic $B\to \pi$ form factor. Combining the inclusive and exclusive results we obtain:
\bea
\left| V_{ub} \right| =
(37.2 \pm 2.7 ) \times
10^{-4} \; .
\eea
which is now in good agreement (within $0.7 \sigma$) with all three central values given above.      

Combining that with the inclusive and exclusive extractions of  $V_{cb}$ we obtain:
\bea
\left| V_{ub} / V_{cb} \right| = 0.0924 \pm 0.0071 \; .
\eea

We now present the result we obtain in the full fit (i.e. including $|V_{ub}|$) in Fig.~\ref{fig:utfit}. The prediction we obtain reads:
\bea
\left[\sin ( 2 \beta) \right]^{\rm prediction}_{\rm full \; fit} = 0.75 \pm 0.04 \; .
\label{sin2beta_fullfit}
\eea
which is consistent, within errors, with Eq.~(\ref{sin2beta_novub}) as
well as with the analyses in Refs.~\cite{Charles:2004jd,Bona:2006ah}.

Fig.~\ref{fig:utfit} clearly emphasizes  a very important role of $V_{ub}$ in that it is very useful in excluding a second solution for the fit to $\sin(2\beta)$ in addition to potentially reducing the errors appreciably (see also Ref.~\cite{Fleischer:2003xx} for a discussion of the second solution).

The deviation from the experimental results are summarized in Table~\ref{tab:sin2beta}. Notice that no significant change in the deviations in $B \to \psi K_s$ and in $B \to (\phi, \eta', K_sK_s)K_s$ has  taken place from the fit that did not use $V_{ub}$.

It is perhaps of some use to extract the values  of $\hat B_K$, $\xi_s$ and $V_{cb}$ that are required to reduce to the 1-$\sigma$ level the discrepancy between the prediction given in Eq.~(\ref{sin2beta_novub}) and  $a_{(\psi+\phi+\eta^\prime+K_S K_S) K_S} = 0.66 \pm 0.024$.    We find that one has to choose either  $\hat B_K^{\rm new} = 0.96 \pm 0.04$,   $\xi_s^{\rm new} = 1.37 \pm 0.06$ or $V_{cb} = (44.3 \pm 0.6) \times 10^{-3}$.  
\begin{table}[t]
\begin{center}
\begin{tabular}{|c|c|c|c|}
\hline
mode & experiment & no $V_{ub}$ &  with $V_{ub}$ \\ \hline
$a_{\psi K_S}$ & $0.681 \pm 0.025 $  & $2.1 \; \sigma$ & $1.7 \; \sigma$ \\
$a_{\phi K_S}$ & $0.39 \pm 0.17 $ & $2.5 \; \sigma$ & $2.1 \; \sigma$ \\
$a_{\eta^\prime K_S}$ & $0.61 \pm 0.07$ & $2.3 \; \sigma$ & $1.8 \; \sigma$ \\
$a_{K_S K_S K_S}$ & $0.58 \pm 0.20$ & $1.4 \; \sigma$ & $0.9 \; \sigma$ \\
$a_{(\phi+\eta^\prime+K_S K_S) K_S}$ & $0.58 \pm 0.06$ & $2.7 \; \sigma$ & $2.5 \; \sigma$ \\
$a_{(\psi+\phi+\eta^\prime+K_S K_S) K_S}$ & $0.66 \pm 0.024$ & $ 2.3\; \sigma$ & $ 2.1\; \sigma$ \\
\hline
\end{tabular}
\caption{Experimental determinations of $\sin (2 \beta)$ in $b\to c\bar c s$ and $b\to s \bar s s$ decays~\cite{Barberio:2007cr}. Also shown are the deviations from the SM prediction obtained without (Eq.~(\ref{sin2beta_novub})), and with,  (Eq.~(\ref{sin2beta_fullfit})) the inclusion of $V_{ub}$ in the fit. \label{tab:sin2beta}}
\end{center}
\end{table}

To summarize, in the SM picture of flavor and CP violation there  seem to be some inconsistencies that might be an important hint of new physics at the electroweak scale. In particular, the predicted value of $\sin (2 \beta)$ in the SM seems to indicate possible inconsistencies with the value directly measured via the gold-plated $B \to \psi K_s$ mode and also by the penguin-dominated modes, such as  $B\to (\phi,\eta^\prime, K_S K_S) K_S$. Furthermore, recall also that it seems rather difficult to reconcile the observed difference~\cite{hfag06} ($\Delta A_{CP}^{exp} = (14.4 \pm 2.9) \times 10^{-3}$) in the direct CP asymmetries of $B^0 \to K^- \pi^+$ and $B^- \to K^- \pi^0$ with our prediction $\Delta A_{CP} = (2.5 \pm 1.4) \times 10^{-3}$~\cite{Lunghi:2007ak} based rather loosely on QCD factorization~\cite{Beneke:2003zv} and obtained by allowing several input parameters to simultaneously take the values at the edge of their range. If these difficulties persist then they would require the need for a beyond the SM CP-odd phase. It does not seem like the problems in the determination of $V_{ub}$ are the cause of these discrepancies; so our earlier conclusions regarding the need for a new-phase are substantiated~\cite{Lunghi:2007ak}. We also note in passing the interesting recent work~\cite{Bona:2008jn} which indicates presence of new physics in their analysis of $B_s-\bar B_s$ mixing (See also Ref.~\cite{Bona:2007vi}). From the perspective of our studies in~\cite{Lunghi:2007ak} and the current work, which suggest the need for a new phase in $b \to s$, the findings of~\cite{Bona:2008jn} appear naturally related. Therefore, there is a heightened need for further clarifications on these important issues. Improved measurements of $\sin (2 \beta)$ via $B \to \psi K_s$ or via the penguin modes, {\it e.g.} $B \to \eta' K_s$, may well have to await new experimental facilities, such as a Super-B (or Super-Flavor) Factory~\cite{Hashimoto:2004sm,Bona:2007qt}. The predicted value of $\sin (2 \beta)$ relies heavily on lattice determinations of $V_{cb}$, the SU(3) breaking ratio $\xi_s$, the kaon parameter $B_K$, and $V_{ub}$. The importance of further improvements in these calculations can hardly be over emphasized.

We thank Andreas Kronfeld, Jack Laiho and Ruth Van de Water for useful discussions. This research was supported in part by the U.S. DOE contract No.DE-AC02-98CH10886(BNL) and in part by the Department of Energy under Grant DE-AC02-76CH030000(Fermilab). Fermilab is operated by Fermi Research Alliance, LLC under Contract No.DE-AC02-07CH11359 with the United States Department of Energy.

\end{document}